# Spin field-effect transistors based on massless birefringent Dirac fermions in polar Dirac semimetals

Hu Zhang*, Chendong Jin, Ruqian Lian, Peng-Lai Gong, RuiNing Wang, JiangLong Wang, and Xing-Qiang Shi

Key Laboratory of Optic-Electronic Information and Materials of Hebei Province, Institute of Life Science and Green Development, College of Physics Science and Technology, Hebei University, Baoding 071002, P. R. China

* E-mails: zhanghu@hbu.edu.cn

The Datta-Das-type spin field-effect transistor, using a two-dimensional electron gas in a semiconductor heterostructure as a channel, plays a key role in spintronics. Here, we theoretically present a type of spin field-effect transistor based on massless birefringent Dirac fermions in polar Dirac semimetals. The manipulation of spin arises from the existence of the strong spin-orbit coupling, polar space groups, and Dirac cones in a single phase. The oscillatory channel conductance can be controlled by the sign of gate voltage in addition to its magnitude due to the gapless band structures of polar Dirac semimetals. Such spin field-effect transistor provides guidance for the further design of spintronic devices.

Datta and Das have proposed the concept of spin field effect transistor (spin FET) in 1990.[1] This spintronic device is based on the precession of the injected spins due to the gate voltage-controlled spin-orbit coupling (SOC) in narrow-gap semiconductors. The spin FET was not realized experimentally until 2009.[2] In the experiment, ferromagnetic electrodes were used to inject and detect spin with a two-dimensional electron gas (2DEG) channel in InAs heterostructures. An oscillatory channel conductance as a function of gate voltage was observed. After that, an all-semiconductor and all-electric spin FET was reported by using quantum point contacts (QPCs) as spin injectors and detectors instead of ferromagnetic components.[3]

An InGaAs heterostructure was used in the experiment. The QPCs can provide spin selection with nearly 100% efficiency. This spin FET only worked well at very low temperature (300 mK). In addition, a spin FET with bulk ferroelectrics GeTe having the giant Rashba effect as a channel was designed.[4] However, there are still many technological challenges for practical applications up to now, especially the low temperature working conditions.[5] It is important to explore other materials and physical mechanism to design spin FETs with high performances at room temperature.[6]

Different from already reported spin FETs using semiconductors as a channel, here we theoretically propose a type of spin FET based on massless birefringent Dirac fermions in polar Dirac semimetals.[7-10] In contrast with the centrosymmetric Dirac semimetals (e.g. $Na_3Bi$),[11] a polar Dirac semimetal have a polar space group in which inversion symmetry is broken.[8] We demonstrate that SOC and the polar character make it possible to use polar Dirac semimetals as a channel. Polar Dirac materials are suitable for device applications since they can have several advantages such as high carrier mobility, Fermi velocity and the easily tunable Fermi level similar to those in centrosymmetric Dirac materials including two-dimensional graphene and three-dimensional $Cd_3As_2$.[12,13]

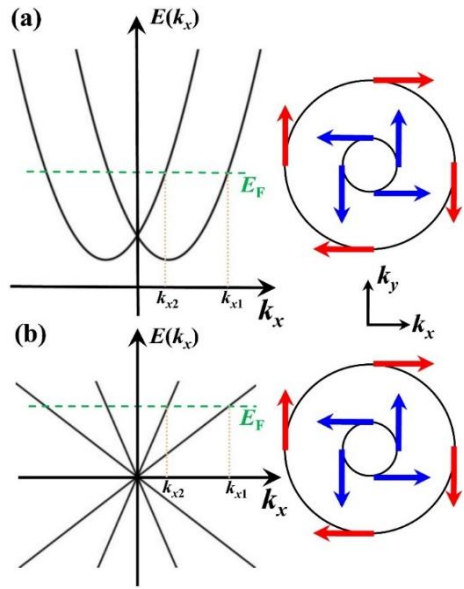

FIG. 1. The energy band structures along the $k_x$ direction and the corresponding Fermi surface and spin textures in the $k_x$-$k_y$ plane at the Fermi level $E_F$ denoted by horizontal dashed lines for (a) the two-dimensional electron gas described by a Rashba term and (b) the massless birefringent Dirac fermion in polar Dirac semimetals. At the Fermi level $E_F$, electrons will have two wave vectors $k_{x1}$ and $k_{x2}$ due to the lock of inversion symmetry in both cases.

The physical mechanism of a Datta-Das type spin FET is the existence of the strong SOC in a 2DEG (in the *x-y* plane) described by a Rashba term in the effective Hamiltonian[1]:

$$H_R(k_{x,y}) = \eta_R(k_x\sigma_y - k_y\sigma_x), \tag{1}$$

where $\eta_R$ is the spin-orbit coefficient, $k_{x,y}$ are wave vectors, and $\sigma_{x,y}$ are the Pauli matrices. The Rashba term can split the energy of spin-up and spin-down electrons. For an electron traveling in the *x* direction ($k_y = 0$ and $k_x \neq 0$), we have

$$H_R(k_x) = \eta_R k_x \sigma_y. \tag{2}$$

The corresponding energy bands along the $k_x$ direction and the spin-textures of the bands at the dashed line indicated energy $E_F$ have been shown in Fig 1(a). We can find that the spin is always perpendicular to the momentum, which is known as the spin-momentum locking effect.[14] The outer and inner bands have clockwise and counterclockwise spin textures, respectively. The energy bands also imply that the $-y$ polarized and $+y$ polarized electrons with the same energy $E_F$ have different wave vectors $k_{x1}$ and $k_{x2}$ due to the Rashba term. Thus, the electrons feel a spin-orbit (SO) effective magnetic field $B_y^{SO}$.[15] This field is usually called the Rashba field, which can lead to the precession of the injected spins. We have

$$E(+y \text{ pol.}) = \frac{\hbar^2 k_{x1}^2}{2m^*} - \eta_R k_{x1} + C, \tag{3a}$$

$$E(-y \text{ pol.}) = \frac{\hbar^2 k_{x2}^2}{2m^*} + \eta_R k_{x2} + C, \tag{3b}$$

where $m^*$ is the effective mass of the electrons and *C* is the constant term. Then we can obtain the difference of wave vectors $k_{x1}$ and $k_{x2}$ for the same energy

$$k_{x1} - k_{x2} = \frac{2m^*\eta_R}{\hbar^2}. \tag{4}$$

On the other hand, the differential phase shift[1] of the electron with different wave vectors $k_1$ and $k_2$ passing through a channel with a length of $L$ is

$$\Delta\theta = (k_1 - k_2)L. \quad (5)$$

For the $-y$ polarized and $+y$ polarized electrons in our case, we have

$$\Delta\theta = (k_{x1} - k_{x2})L = \frac{2m^*\eta_R L}{\hbar^2}. \quad (6)$$

The differential phase shift is proportional to $\eta_R$ which can be modified by a gate voltage experimentally. This results in an oscillatory channel conductance with respect to the change of the applied gate voltage. These results obtained by Datta and Das are the basic physical mechanism of a spin FET using a 2DEG in semiconductor heterostructures.

Now we investigate the physical mechanism of the spin FET using massless birefringent Dirac fermions in polar Dirac semimetals. Na$_3$Bi with centrosymmetric *P*6$_3$/*mmc* symmetry is the firstly reported three-dimensional Dirac semimetal whose conduction and valence bands touching at the Fermi energy.[11] Such rotational symmetry protected touch point with linear energy dispersion is called the Dirac point. Around the Dirac point, the low energy physics can be described by massless Dirac fermions.[13] There exists another type of three-dimensional Dirac semimetal with a polar space group in which inversion symmetry is broken. Previously reported works have found Dirac points in CaAgAs with the *P*6$_3$*mc* polar space group.[7,8] Different from Na$_3$Bi, energy bands around such Dirac points that do not along the rotational axis are now split by SOC as a result of the lack of inversion symmetry, which will be discussed below. These types of Dirac semimetals can be called *polar* Dirac semimetals to exhibit their ferroelectric-like polar symmetry. Consequently, Dirac fermions in these materials are known as massless birefringent Dirac fermions.[9]

In our previous work,[8] based on the theory of invariants, we have found that the *k·p* low energy effective Hamiltonian (up to linear order in *k*) for the massless birefringent Dirac fermions in polar Dirac semimetals (the rotational axis is along the *z* direction) can be described by

$$H_{BDF}(k) = \begin{pmatrix} h_1(k) & 0 \\ 0 & h_2(k) \end{pmatrix}, \quad (7)$$

with $h_1(k) = \alpha_1 k_z \sigma_0 + \eta_1(k_x \sigma_y - k_y \sigma_x)$, $h_2(k) = \alpha_2 k_z \sigma_0 + \eta_2(k_x \sigma_y - k_y \sigma_x)$, $\sigma_0$ is the identity matrix, $\alpha_{1,2}$ and $\eta_{1,2}$ are materials-dependent model parameters which are all real, and BDF is short for the birefringent Dirac fermion. This effective Hamiltonian is significantly different from that of Dirac fermions in centrosymmetric Dirac semimetals.[13] The band dispersion for the wave vector $k_z$ is $E(k_z) = \alpha_{1,2} k_z$. For wave vectors in the $k_z = 0$ plane, we have

$$H_{BDF}(k_{x,y}) = \begin{pmatrix} \eta_1(k_x \sigma_y - k_y \sigma_x) & 0 \\ 0 & \eta_2(k_x \sigma_y - k_y \sigma_x) \end{pmatrix}. \tag{8}$$

Compared with the Rashba term $H_R$ in Eq. (1), one can find that $H_{BDF}$ is actually the double version of $H_R$. This is the key physical feature of the birefringent Dirac fermions considered here.

The band dispersion for the wave vector $k_{x,y}$ around Dirac points is

$$E(k_{x,y}) = \pm \eta_{1,2} \sqrt{k_x^2 + k_y^2}. \tag{9}$$

In the $k_z = 0$ plane, the Fermi velocity around Dirac points has two different values $\eta_1$ and $\eta_2$ due to the lack of inversion symmetry. For an electron traveling in the $x$ direction ($k_y = 0$ and $k_x \neq 0$), we have

$$E(k_x) = \pm \eta_{1,2} k_x. \tag{10}$$

The corresponding energy bands along the $k_x$ direction and spin-textures of bands at energy $E_F$ have been shown in Fig 1(b). In this case, the spin is also always perpendicular to the momentum, showing a spin-momentum locking effect. The outer and inner bands have clockwise and counterclockwise spin textures respectively. The energy bands indicate that the $-y$ polarized and $+y$ polarized electrons with the same energy $E_F$ now have different wave vectors $k_{x1}$ and $k_{x2}$ due to SOC in polar Dirac semimetals. The electrons also feel an effective magnetic field $B_y^{SO}$. At $E_F$, we have

$$\eta_1 k_{x1} = E_F = \eta_2 k_{x2}. \tag{11}$$

Then we can obtain the difference of wave vectors $k_{x1}$ and $k_{x2}$ for the same energy

$$k_{x1} - k_{x2} = \frac{E_F}{\eta_1} - \frac{E_F}{\eta_2} = (\frac{1}{\eta_1} - \frac{1}{\eta_2}) E_F. \tag{12}$$

Thus, the differential phase shift of the $-y$ polarized and $+y$ polarized electrons passing through a channel with a length of $L$ is

$$\Delta\theta_{BDF} = (k_{x1} - k_{x2})L = \left(\frac{1}{\eta_1} - \frac{1}{\eta_2}\right)E_F L. \qquad (13)$$

Now the differential phase shift is related to $L$, Fermi level $E_F$, and $\eta_{1,2}$. In a 2DEG, the differential phase shift is independent of the Fermi level, as shown in Eq. (6). This is a very important difference between spin FETs using a semiconductor heterostructure channel and a polar Dirac semimetal channel. This character allows the controlling of the differential phase shift by tunning the Fermi level. The Fermi level $E_F$ is considered to be easily tunable by a gate voltage as in the case of graphene.[16] The Fermi velocity $\eta_1$ and $\eta_2$ are related to the spin-orbit coefficient and thus can be modified by a gate voltage experimentally. As a result, the differential phase shift of the polarized electrons passing through a polar Dirac semimetal channel can be controlled by a gate voltage.

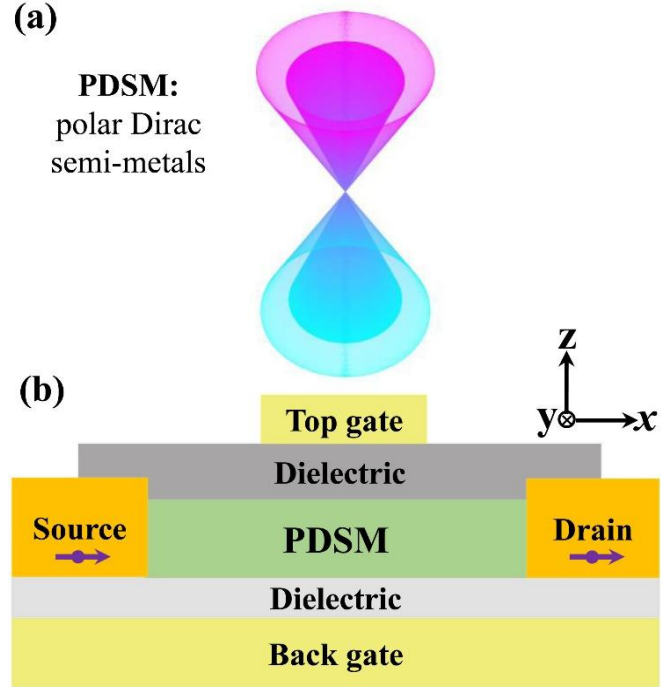

FIG. 2. (a) The Dirac cones in the polar Dirac semi-metals (PDSM). (b) The basic structure of a spin field effect transistor based on the polar Dirac semimetals.

Based on the above discussions, we now investigate spin FETs using a polar Dirac semimetal as a channel instead of a 2DEG formed in semiconductor heterostructures. In Fig. 2(a), we have plotted the Dirac cones in polar Dirac semimetals. The linear energy dispersion and its splitting due to the breaking of the inversion symmetry are clearly shown. The basic structure of the device based on the polar Dirac semimetal is given in Fig. 2(b). We can use ferromagnetic electrodes or QPCs as the source and drain to inject and detect spin respectively. The injected spin will precess when passing through the polar Dirac semimetal channel in the $x$ direction under an effective magnetic field $B_y^{SO}$. The differential phase shift $\Delta\theta_{BDF}$ given in Eq. (13) can be controlled by the gate and thus can vary from 0 to $2\pi$. The detector voltage will be low when a detected spin has its orientation antiparallel (with a phase difference of $\pi$) with that of the detector and will be high when the spin is parallel. As a result, the channel conductance will oscillate periodically as a function of increasing gate voltage. This kind of spin FET will benefit from the good physical properties of polar Dirac semimetals including high carrier mobility and Fermi velocity.

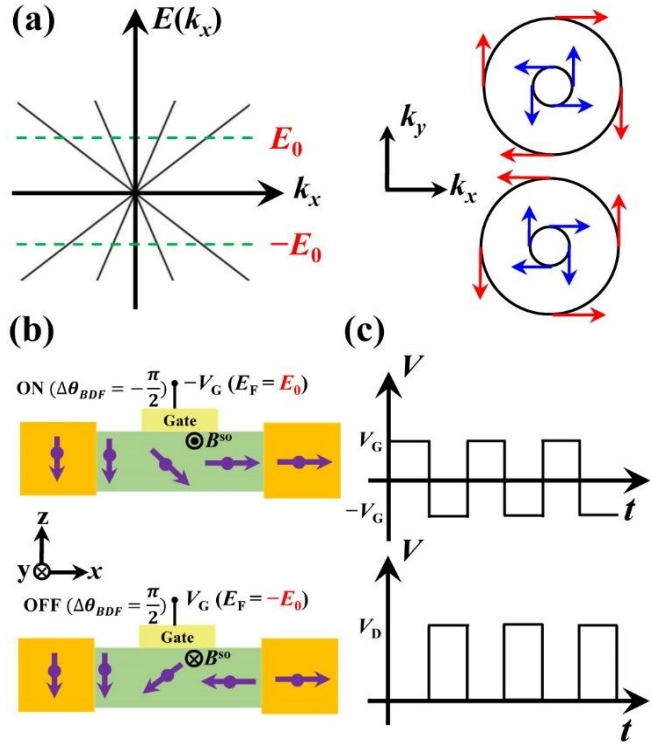

FIG. 3. (a) The energy band structures of the massless birefringent Dirac fermion in polar Dirac semimetals along the $k_x$ direction and the corresponding Fermi surface and spin textures in the $k_x$-$k_y$ plane at the Fermi level $E_0$ ($E_0 > 0$) and $-E_0$ denoted by horizontal dashed lines. (b) The alteration of the sign of the differential phase shift ($\Delta\theta_{BDF}$) with the reverse of the direction of an effective magnetic field $B^{SO}$ by changing the sign of applied gate voltage ($V_G$). The magnetization of the injector and detector are oriented along the *z* and *x* axis respectively. (c) Schematic plot of the relation between the applied gate voltage and the detector voltage ($V_D$).

The unique gapless band structures of the polar Dirac semimetal also make it possible to control the *sign* of the differential phase shift by adjusting the position of Fermi level of the polar Dirac semimetal. As shown in Fig. 3(a), the outer and inner bands have clockwise and counterclockwise spin textures, respectively, when the Fermi level is tuned to $E_0$ (> 0). And if the Fermi level is tuned to $-E_0$, the outer and inner bands have counterclockwise and clockwise spin textures respectively. This results in a reverse of the direction of an effective magnetic field $B^{SO}$ felt by electrons traveling in polar Dirac semimetals. Therefore, we can alter the sign of the differential phase shift by changing the sign of the applied gate voltage.

We have illustrated the detailed physical mechanism in Fig. 3(b). The magnetization of the injector and detector are oriented along the *z* and *x* axis respectively. The injected spins are polarized along the negative *z* direction and will process under the influence of $B^{SO}$ traveling in the polar Dirac semimetal channel. We can apply a gate voltage $-V_G$ ($V_G$ > 0) to tune the Fermi level $E_F$ to $E_0$. In this case, the effective magnetic field $B^{SO}$ is along the negative *y* direction and thus the injected spin could rotate counterclockwise. According to the relation given in Eq. (13), a phase shift of $-\pi/2$ can be obtained by choosing a suitable gate voltage $-V_G$. The injected electrons can pass through the detector since their spin rotates to be parallel to the polarization direction of the detector. We will have an "ON" state. On the other hand, when we apply a gate voltage $V_G$ ($V_G$ > 0) to tune the Fermi level $E_F$ to $-E_0$, the effective magnetic field $B^{SO}$ is now along the positive *y* direction. The injected spin now rotates clockwise. In principle, we can obtain a phase shift of $\pi/2$. Now, the injected electrons cannot pass through the detector since their spin rotates to be antiparallel to the polarization direction. We can have an "OFF" state. When an oscillated gate voltage in rectangular pulses is applied as shown in Fig. 3(c), the output detector voltage presents the same frequency as the gate voltage. This is an advantage for spin FET applications.

In conclusion, we have proposed a type of spin FET using massless birefringent Dirac fermions in polar Dirac semimetals. The low energy physics of these quasi-particles is described by an effective Hamiltonian that can be viewed as the double version of the Rashba term. The intertwined SOC, polar symmetry, and Dirac points make it possible to use polar Dirac semimetals as a channel. This spin FET would benefit from the common outstanding physical properties of Dirac materials such as high carrier mobility and Fermi velocity. Furthermore, the manipulation of spin degrees of freedom discovered here offers the possibility of designing other spintronic devices using polar Dirac semimetals.

This work was supported by the Advanced Talents Incubation Program of the Hebei University (Grants No. 521000981423, No. 521000981394, No. 521000981395, and No. 521000981390), the Natural Science Foundation of Hebei Province of China (Grants No. A2021201001 and No. A2021201008), the National Natural Science Foundation of China (Grants No. 12104124, No. 11904154, and No. 51772297), and the high-performance computing center of Hebei University.

**AUTHOR DECLARATIONS**
**Conflict of Interest**
The authors have no conflicts to disclose